\documentclass[]{spie}  
\pdfoutput=1
 
\usepackage{amsmath,amsfonts,amssymb}
\usepackage{graphicx}
\usepackage[colorlinks=true, allcolors=blue]{hyperref}

\title{A first measurement of the Planetary Boundary Layer top in Cali-Colombia: Elastic LiDAR application}

\author[a,b]{Jonnathan C\'espedes}
\author[a,c]{Carlos Andr\'es Melo-Luna}
\author[a,c]{John H. Reina}
\affil[a]{Centre for Bioinformatics and Photonics (CIBioFi), Calle 13 No. 100-00, Edificio E20 No. 1069, 760032 Cali, Colombia}
\affil[b]{Escuela de Ingenier\'ia de los Recursos Naturales y el Ambiente (EIDENAR), 760032 Cali, Colombia}
\affil[c]{Departamento de F\'isica, Universidad del Valle, 760032 Cali, Colombia}

\authorinfo{Further author information: (Send correspondence to Jonnathan Cespedes)\\E-mail: jonnathan.cespedes@correounivalle.edu.co, Telephone: +57 3167499451}

\begin{document} 
\maketitle

\begin{abstract}
The monitoring of the impact of aerosols in Latin America on a local scale is usually limited due to the infrastructure and instrumentation available. In Colombia, there are two international ground surface monitoring networks, the AErosol RObotic NETwork (AERONET) and the Latin American LIDAR NETwork (LALINET). However, the AERONET performance relies on only one sun photometer which makes measurements distributed among five ground-based stations in different cities such as Bogot\'a and Medell\'in. On the other hand, LALINET has only one ground-based station formed by an elastic LiDAR system located at Medell\'in. Although Cali is the largest city of Colombian southwestern, with an accelerated grown rate of both urban and vehicular fleet, and counts with the third largest population of this country, is not reached by these networks. Here, we report on the implementation of a monostatic-coaxial multispectral LiDAR system using a pulsed Nd:YAG laser with 450 mJ of average energy at 1064 nm. To perform the atmospheric measurements, this system is capable of spatially resolving elastic backscatter down to 3.75 m with a Pulse Repetition Frequency (PRF) of 10 Hz. We have developed a hybrid algorithm for data analysis by combining the Fitting and Gradient method and the Klett-Fernald algorithm to estimate the Planetary Boundary Layer (PBL) Top and the optical properties of aerosols. This work constitutes the first quantitative atmospheric exploration to study the aerosols dynamics and the PBL in the northwest of South America.
\end{abstract}


\section{INTRODUCTION}
\label{sec:intro}  

\begin{figure} [ht]
   \begin{center}
   \begin{tabular}{c} 
   \includegraphics[width=23cm,height=13cm,keepaspectratio]{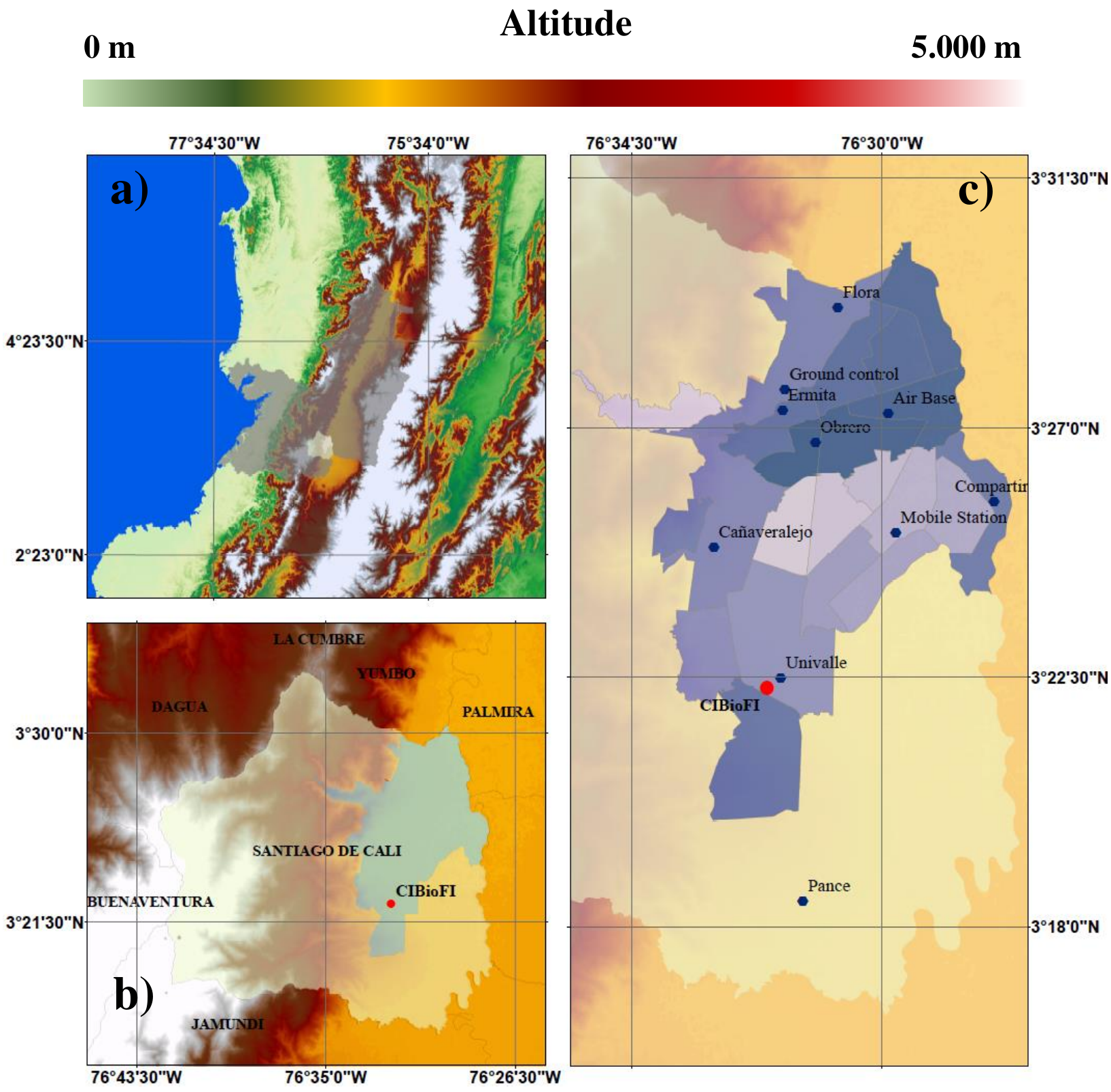}
   \end{tabular}
   \end{center}
   \caption[example] 
   { \label{fig:ze} 
{\bfseries a)} The general location of Valle del Cauca department (Gray area) and the municipality of Cali (White area) within Colombia, with altitude values according to the color scale. {\bfseries b)} The light color shape represents the rural area of Cali whereas the Blue area is related to the urban area of the city. The Red dot mark the location of the LiDAR-CIBioFI system at Universidad del Valle. {\bfseries c)} Political and administrative division of Cali, the Blue dots locate the governmental ground stations of quality air programmes (SVCA-``Sistema de Vigilancia de Calidad del Aire'' of DAGMA-``Departamento Administrativo de Gesti\'on del Medio Ambiente'').}
   \end{figure}

Located in Colombia-South America, Santiago de Cali (Cali, for short, $3^\circ 27'00''N; \ 76^\circ 32'00''W)$ is the capital of the Valle del Cauca department. It is limited to the north by Yumbo and La Cumbre municipalities, to Northeast by Palmira, to East by Candelaria, to Southwest by special economic district Buenaventura, and to Northwest by Dagua. The west and central Andean Mountains Ranges define the geography of this city in which the urban region is mainly located between flat topography and low hills (See Fig.~\ref{fig:ze}). Its location has a mean altitude of 1.000 m above the sea level with a tropical savanna climate, an average temperature of around $25^{o}C$, and annual precipitations that reach up to 5.000 mm.  In spite that the west mountain chain blocks most of the Sea breeze from the Pacific Ocean, some sea wind overpasses the natural barrier decreasing the high temperatures of this city. This latest issue and the elevation variations in the mountain-chain (from 2000 m to 4100 m, from north to south respectively) provide rainfalls in the city south-west.

Cali had a population of about  2.075.380 people in 2005 according to the last population study by the National Administrative Department of Statistics (DANE by its Spanish meaning, Departamento Administrativo Nacional de Estad\'istica).  Currently, there is an estimation of around 2.800.000 inhabitants in 2018. This city is decisive to the country's economy because of the quick connection to the sea through Buenaventura seaport and to the industrial district of Yumbo. Buenaventura is the primary  colombian seaport and moves about 60\% of the merchandise that arrives and departs from the country. On the other hand, Yumbo industrial park is a fundamental part of the national development due to the manufactured preponderance of consumer goods. 

The commodities transported from Buenaventura also mean a negative impact on the local air quality due to the additional amount of substances loaded into the atmosphere, besides the local vehicular emissions which have increased around 30\% since 2002, counting with 671.607 gas-powered vehicles at the end of 2017 \cite{fleet}. However, the real effects of vehicular emissions on the air quality of Cali have not been studied in detail \cite{silva13} up to now. On the other hand, the pollutant emissions by the industrial park Yumbo have been the focus of most studies \cite{jaramillo04,filigrana2012blood,jime2011} because of the proximity to Cali and the evidence that these emissions have a measurable influence over public health due to wind-related dynamics of aerosols and trace gases. Nevertheless, these studies have not been rigorously considered by the local administration on the definition of public policies for the protection of air quality.

In Cali operates the Air Quality Monitoring System - SVCA (by its meaning Spanish), run by the Administrative Department of Environmental Management - DAGMA (by its  Spanish meaning) \cite{dagma}. The SVCA is composed of nine ground stations as shown in Fig.~\ref{fig:ze}, and it measures the concentration to an urban receptor altitude level of $PM_{10}$, $PM_{2.5}$, $SO_{2}$, $NO_{2}$, $O_{3}$ and $H_{2}S$. Moreover, the SVCA issues a monthly report indicating the allowed levels of the mentioned substances contrasted to local regulations (Resolution 610 of 2010 \cite{res610} and 1541 of 2013\cite{res1541}), and international standards from the Environmental Protection Agency - EPA\cite{epaindex}.

In spite of existing the  SVCA that measures $PM_{10}$  and $PM_{2.5}$, there are no reports related either to aerosols dynamics in the vertical atmospheric column or to the local Planetary Boundary Layer (PBL) due to the unavailability of stations from AERONET\cite{holben1998aeronet} or LALINET\cite{guerrero2016latin}. To account for this, the Centre for Bioinformatics and Photonics (CIBioFI) at Universidad del Valle has developed a multispectral LiDAR system to support the monitoring of aerosols profiles, their optical properties, and the PBL. We collect the data and analyze it in two phases: a) using the Klett-Fernald algorithm to calculate the optical properties like the extinction and backscatter coefficients\cite{klett1981stable, klett1985lidar}, b) employing a hybrid algorithm of gradient and Wavelet Covariance Transform (WCT) methods to detect the top of the PBL. This paper presents the first experimental attempt quantify the vertical atmospheric column in Cali and the Colombian southwestern reporting initial estimations of PBL top and aerosol profiling to understand the role of aerosols in the local climate system.


\section{EXPERIMENTAL METHODS}

\subsection{LiDAR-CIBioFI Instrumentation}
\label{sec:instru}
\begin{figure} [ht]
   \begin{center}
   \begin{tabular}{c} 
   \includegraphics[width=9.8cm,height=9.8cm,keepaspectratio]{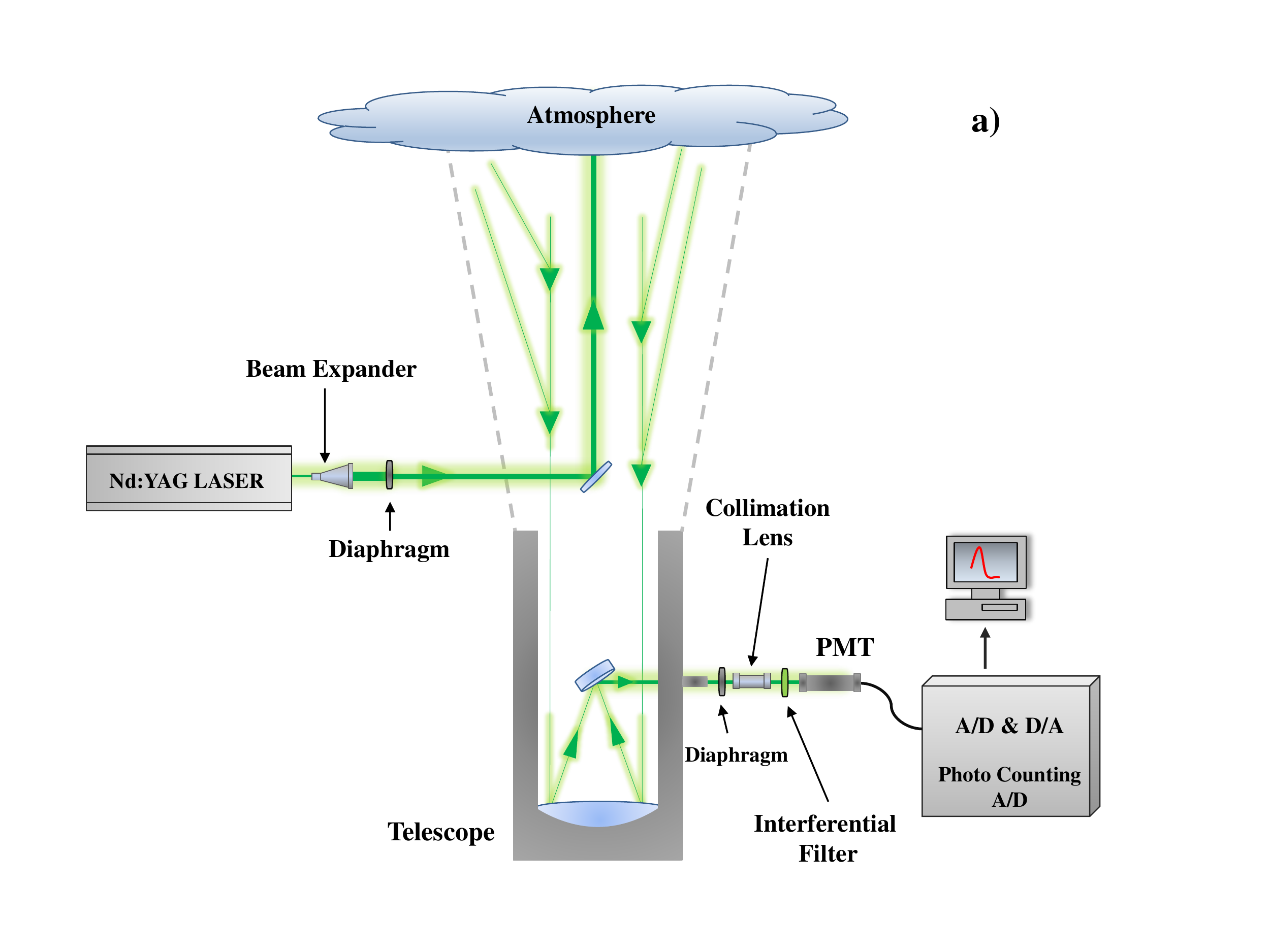}
   \includegraphics[width=8cm,height=8cm,keepaspectratio]{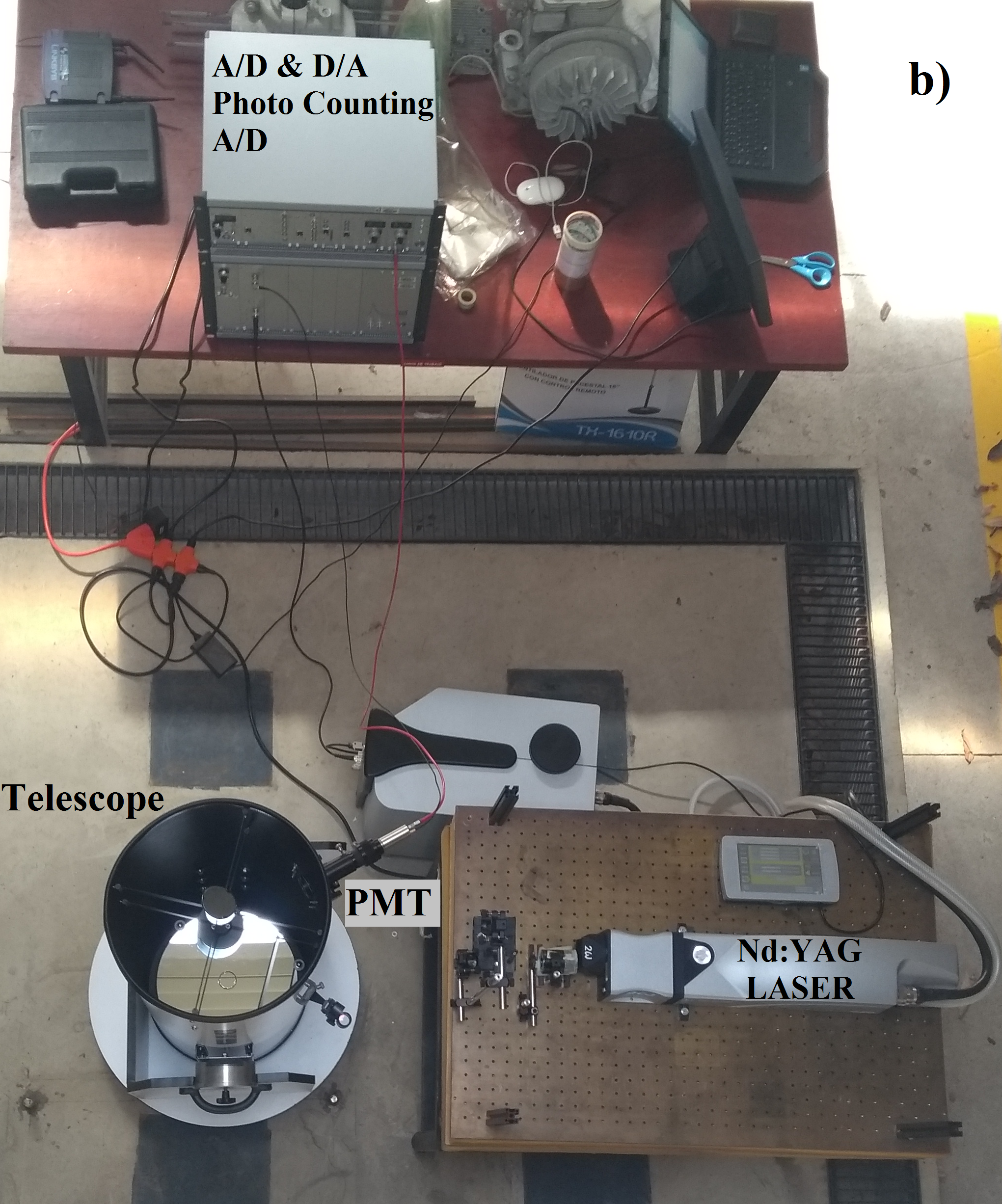}
   \end{tabular}
   \end{center}
   \caption[example] 
   { \label{fig:pic} 
{\bfseries a)} The LiDAR-CIBioFI  setup configured in a coaxial statical mode at 532 nm of the emission wavelength. {\bfseries b)} Topview of the LiDAR-CIBioFI in situ.}
   \end{figure}

We refer to our LiDAR system as LiDAR-CIBioFI, which is located at Universidad del Valle $(3.37 N; 76.53 W)$, Cali- Colombia. Here, we currently operate the LiDAR-CIBioFI in the elastic mode to receive single-scatter signals at 532 nm of wavelength and with a coaxial static alignment configurated as shown in Fig.~\ref{fig:pic}.

The emission component comprises as a light source the single frequency Nd: YAG pulsed laser (Quantel model Q-smart 450 mJ) with a pulse width $\leq$ 6 ns at 1064 nm, and a PRF of 10 Hz. We use the second harmonic generator (SHG) to double the fundamental frequency resulting in a wavelength of 532 nm with an average energy $\geq$ 220 mJ per pulse. The output beam has a diameter of $6.5$ mm which is expanded (3x) three times through a Galilean system with a divergence lower than $0.5$ mrad and a 99\% of transmittance at 532 and 1064 nm.   

A Newtonian telescope with a focal length of 1 m and a primary mirror with a diameter of $0.3$ m constitutes the reception component.   The telescope with a field view of 1.47 mrad collects the backscattered light by the atmosphere. The collimation lens is coupled to a diaphragm with variable aperture, then to an interferential filter at 532 nm, and finally to a photomultiplier tube (PMT, Hamamatsu, R9880U series) to guarantee the reception of the light elastically scattered with a quantum efficiency of 50\% at 532 nm.   

The transient recorder system (Licel, TR20-160) comprises the detection part. The transient system allows the discrimination of the signal from different altitudes through the synchronization with a periodic signal (the trigger)  which usually comes from the laser and increases a counter at detecting a photonic event, and achieve, in our case, a spatial resolution of 3.75 m in the vertical atmospheric column.  

The overlap function reaches a minimum altitude (at $\sim50$ m) for the signal reception which allows the detection of low PBLs because of the coaxial static configuration of the LiDAR-CIBioFI. Even more, this full overlap allows us the profiling of optical properties of aerosols in low altitudes. In this work, we study the PBL evolution in a  day-long period. The acquisition protocol consists of routinary observations on different days storing LIDAR signals with at least 200 pulses per data since 8:00 until 18:00 in local time.

\subsection{LiDAR equation}
\label{sec:lieq}

The retrieved signal from the elastic-LiDAR due to single-scattering in the zenith direction can be modeled by \cite{kovalev2004elastic}:

\begin{equation}
\label{eq:lidareq}
P(z)=\frac{C\times O(z)}{z^{2}}\beta(z)\exp\bigg(-2 \int_{0}^{r} \alpha(z)dz\bigg) ,
\end{equation}

\noindent where $P(z)$ is the so-called lidar signal which is a function of the altitude $z$, $C$ is the LiDAR constant that depends on the quantum efficiency of the PMT, the  diameter of the telescope primary mirror, the speed of light, the radiative flux of the output laser beam,  and $O(z)$ is the overlap function. $\beta(z)$ and $\alpha(z)$ are the backscatter and extinction coefficients, respectively, considering both the molecular and the aerosols contributions. The divergences in the path introduce modifications over the signal which can be compensated through the logarithmic Range Corrected Signal (RCS) and  allow to obtain a stratification of the atmospheric column. The logarithmic RCS is  given by:

\begin{equation}
\label{eq:rcseq}
x(z)=\ln(P(z)\times z^{2}).
\end{equation}

Due to the fact that the well-mixed layer of the PBL contains the most meaningful aerosols load, it is expected that $x(z)$ within such layer exhibits the highest value of backscattering up to entrainment zone, where the load of aerosols begins to decrease. Hereafter, we will apply some analytical methods to detect the PBL top by the identification of sudden changes in $x(z)$, which are produced by differences in the load of aerosols through the atmospheric column.

\subsection{PBL top estimation}
\label{sec:meth}

Our results are based on the interplay of two well-known methods to retrieve the PBL top. For the sake of completeness, we present first a concise introduction to both the gradient and the wavelet covariance transform methods.

{\bfseries The gradient method} considers that backscattering is stronger in the well-mixed PBL than in the entrainment zone and free atmosphere. Therefore, the signal $x(z)$ should decrease with the altitude, and the derivative of $x(z)$ ($dx(z)/dz$)  exhibits a local minimun in the inflection interval which represents the PBL top. On the other hand, an important feature about this method relies on the presence of several local minima by the  increase of values in $x(z)$ due to the presence of water vapor or aerosol clouds in the recorded LiDAR signal.

{\bfseries The Wavelet Covariance Transform (WCT)} employs the Haar function, which due to its discontinuous behavior, allows the estimation of sudden changes in the signal, e.g., the step changes in $x(z)$. WCT is defined by\cite{brooks2003finding}:

\begin{equation}
\label{eq:wcteq}
WCT(a,b) = \frac{1}{a}\int_{z_{b}}^{z_{t}} x(z) h\bigg(\frac{z-b}{a}\bigg)dz,
\end{equation}

\noindent where $z$ is the altitude, $z_{b}$ and $z_{t}$ are the bottom and top altitudes in the RCS, $a$ and $b$ are the spatial dilation and translation of the central position of the scaling of Haar wavelet $h\big(\frac{z-b}{a}\big)$, and

\begin{equation}
\label{eq:haareq}
h\bigg(\frac{z-b}{a}\bigg) = \left\{
                                  \begin{array}{ccc}
                                   1 & \quad b-\frac{a}{2} \leq z \leq b,\\
                                   -1 & \quad b \leq z \leq b+\frac{a}{2}, \\
                                   0 & \emph{{\normalfont elsewhere.}}
                                  \end{array}
                                \right.
\end{equation}

\begin{figure} [ht]
   \begin{center}
   \begin{tabular}{c} 
   \includegraphics[width=16cm,height=16cm,keepaspectratio]{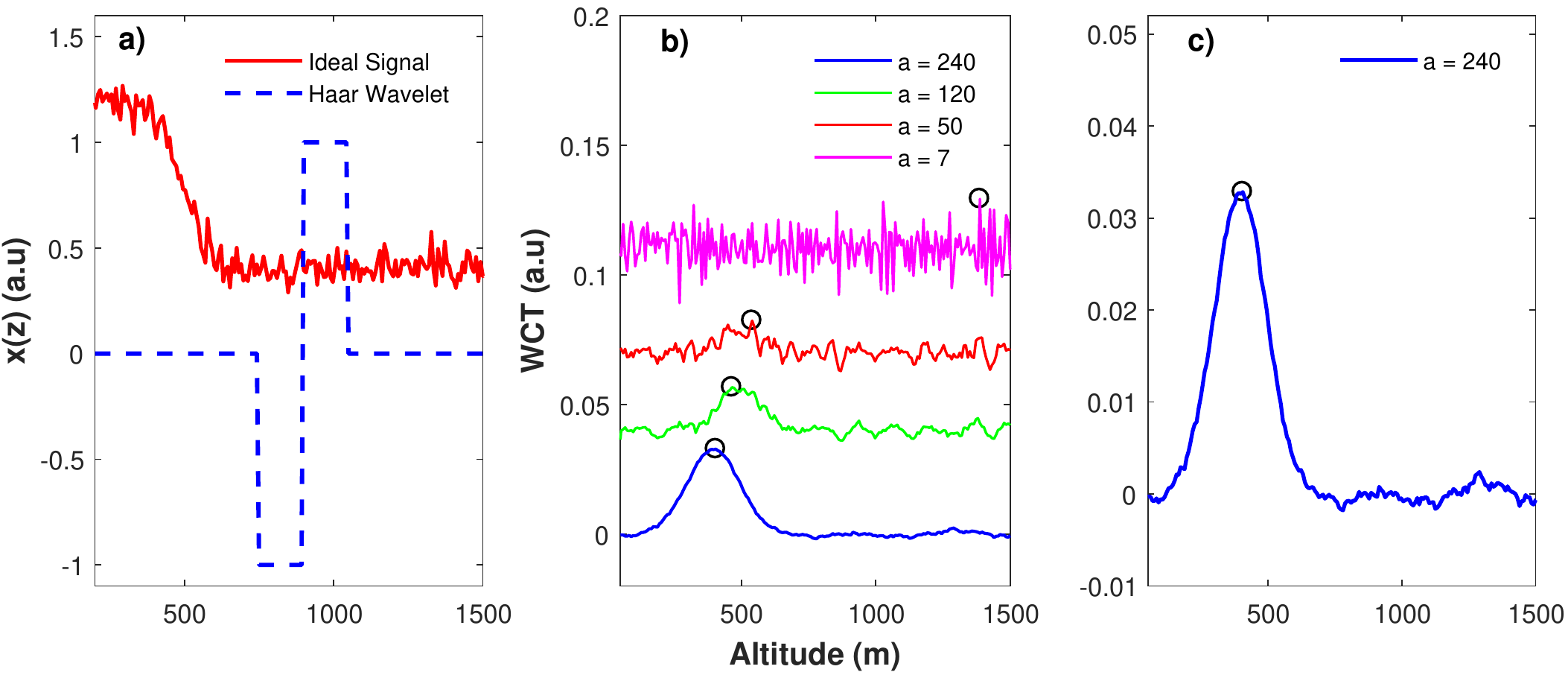}
   \end{tabular}
   \end{center}
   \caption[example] 
   { \label{fig:wctpr} 
{\bfseries a)}  An idealized backscatter lidar profile contaminated with Gaussian noise (Red curve) and Haar wavelet (Blue dashed line). {\bfseries b)} WCTs in dilation $a=240$ (Blue Curve), $a=120$ (Green Curve), $a=50$ (Red Curve), $a=7$ (Magenta Curve), and their PBL like local maximum (Black circles) and {\bfseries c)} WCT (blue curve) for the dilation $a= 240 = 486 \ m$ and the PBL like local maximum (Black circle).}
   \end{figure}

The WCT indicates similarity between $x(z)$ and $h\big(\frac{z-b}{a}\big)$. The Figure~\ref{fig:wctpr}a shows an idealized backscatter profile simulated and modified by a Gaussian noise (standard deviation of 0.05) ensuring a proper noise-signal relation.  Complementary, the Fig.~\ref{fig:wctpr}b shows the application of the WCT to the idealized profiles for different values of $a$, the spatial dilation parameter. One can observe the influence of $a$ on the retrieved WCT results remarking the importance of the appropriate parameter choice.

As shown in Fig.~\ref{fig:wctpr}b, low values for the spatial dilation such as $a=7$ or $a=50$ can also drastically change the idealized backscatter profile or experimental RCS. However, it is possible to observe that the most prominent change using $a=7$ or even $a=50$ is the noise predominance without clarity about what peak represents the PBL top, due to the fact that the results using WCT with small dilations can be similar to the results applying the gradient method over the idealized backscatter profile or experimental RCS \cite{comeron2013wavelet}. Hence, it is not advised to use low values of $a$ for PBL top detection. On the other hand, using higher values such as $a=240$ or $a=120$, the functional form of idealized profiles changes significantly highlighting a peak (local maximum) which represents the PBL top, see Fig~\ref{fig:wctpr}c. Next, we present a practical proposal to ensure an adequate selection of $a$ and $b$ parameters that optimizes the employment of both methods discussed above .

\subsection{Simulated signals}
\label{sec:simu}
   
To evaluate the performance of the methods, we elaborated a dataset with simulated signals under many scenarios for backscattering by the following expression \cite{steyn1999detection}, 

\begin{equation}
\label{eq:simueq}
u(z)=\frac{x_{m}+x_{u}}{2}-\frac{x_{m}-x_{u}}{2}erf(\frac{z-z_{m}}{s}),
\end{equation}

\noindent where $u(z)$ is the ideal simulated backsttering profile which depends on the altitude $z$, $x_{m}$ is the mean of $u(z)$ in the mixed layer, $x_{u}$ is the mean of $u(z)$ in the free troposphere, $z_{m}$ is the mean of PBL depth and $s$ is the entrainment zone thickness. 

\begin{figure} [ht]
   \begin{center}
   \begin{tabular}{c} 
   \includegraphics[width=11cm,height=11cm,keepaspectratio]{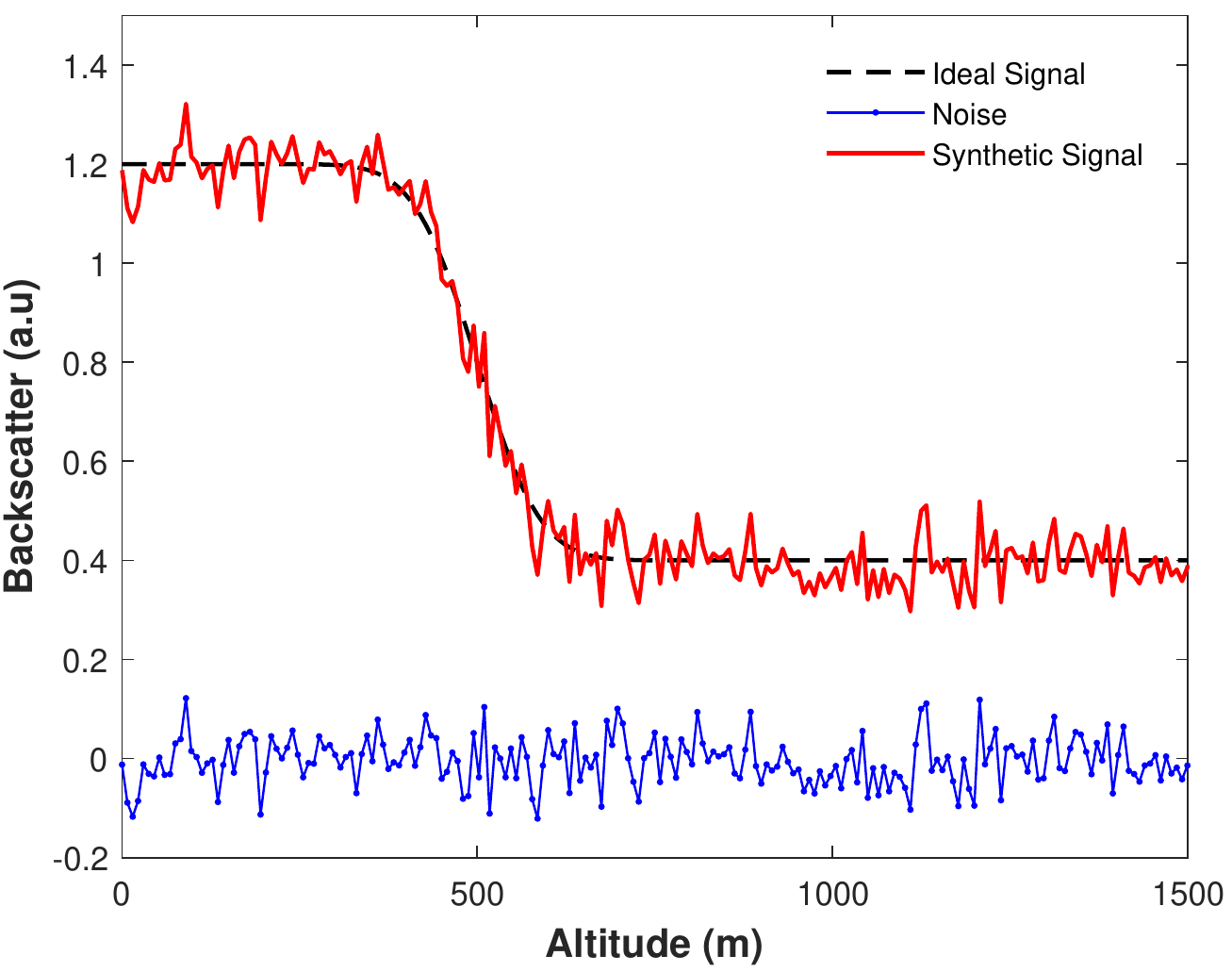}
   \end{tabular}
   \end{center}
   \caption[example] 
   { \label{fig:typ}  A typical scheme of an ideal simulated RCS with its main components.}
   \end{figure}

We simulated two hundred (200) signals using the random parameters of Eq.~\ref{eq:simueq} with large differences between $x_{m}$ and $x_{u}$ to distinguish the well-mixed layer from the free troposphere and by employing several altitudes averages for the entrainment zone. Each signal has an altitude interval ranging from the ground level up to 4 km high. Additionally, each simulated signal was altered by a Gaussian noise (0.05 of standard deviation) with negligible mean value to guarantee a simulated noise smaller than the experimental signals, as showing Figure~\ref{fig:typ}.

\begin{figure} [ht]
   \begin{center}
   \begin{tabular}{c} 
   \includegraphics[width=15.5cm,height=15.5cm,keepaspectratio]{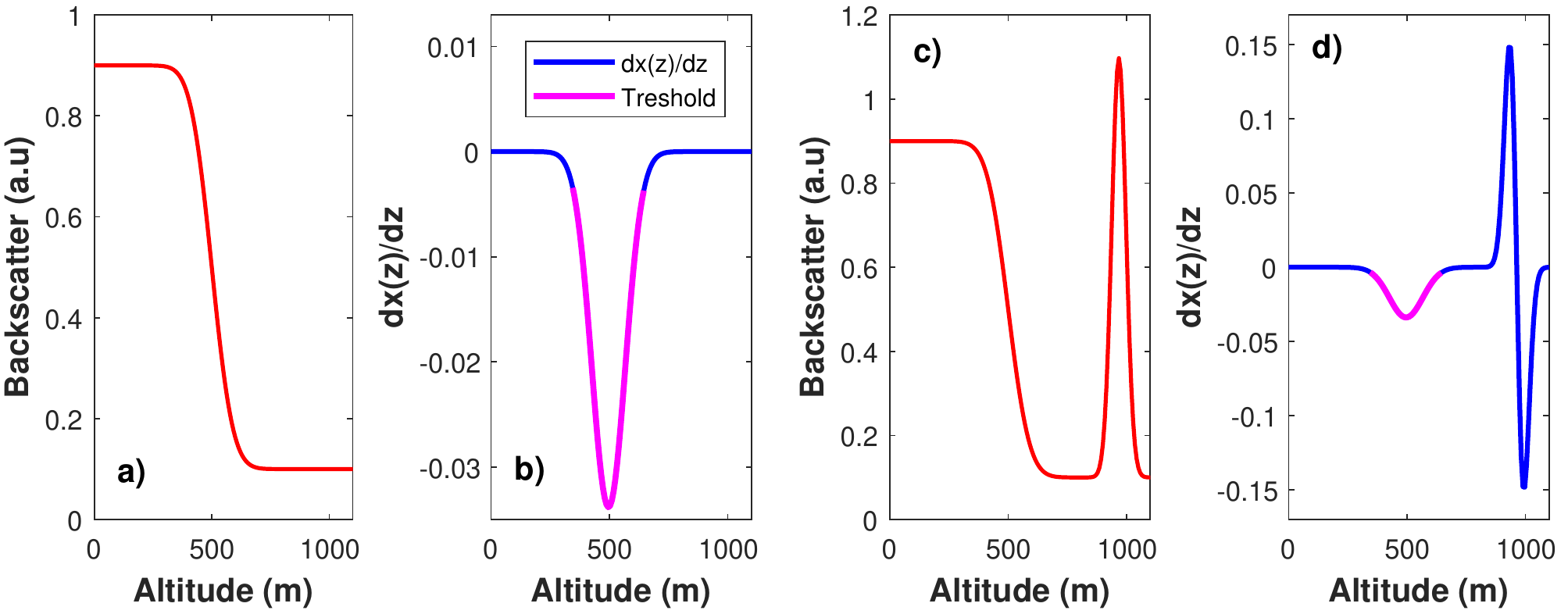}
   \end{tabular}
   \end{center}
   \caption[example] 
   { \label{fig:ideals} 
{\bfseries a)} Under an ideal backscatter profile without noise (Red curve), {\bfseries b)} the Gradient method is applied (blue curve) to obtain a global minimum, and only one inflection interval defines the threshold (Magenta highlight) for the PBL top detection, {\bfseries c)} by including in {\bfseries a)} an additional peak in backscattering due to the cloud presence, {\bfseries d)} now the method displays two extreme minima regarded to entrainment zone and the cloud, the Magenta curve highlights the threshold by the PBL detection in the entrainment zone.}
   \end{figure}

The Gradient method is highly susceptible to produce inappropriate results on noisy signals because the derivative $dx(z)/dz$ may have several local maxima values due to the noise fluctuations. In this sense, we use simulated signals without additional noise to avoid the unreliable results generated by this method as shown in Fig~\ref{fig:ideals}b. Furthermore, if the signal exhibits an increase in backscattering due to cloud presence (Fig.~\ref{fig:ideals}c), the obtained results through the gradient method will have at least two minima, as Fig.~\ref{fig:ideals}d shows.

By contrast, the WCT is less sensitive to the noise presence and offers better results than the gradient method on noisy signals. In this way, we evaluate additional considerations with an experimental signal from the LiDAR-CIBioFI, which we discuss in the following section.
   
\subsection{Gradient method as a threshold for WCT}
\label{sec:grad}

The PBL top does not have a fixed altitude and is one of the most dynamical troposphere variables in climatology since it presents a response to variables such as temperature, precipitations, the topography, and the additional sources generated by human activity. As a consequence, the PBL top behavior pattern requires the constant monitoring and the application of methods for its detection and analysis.  

In Fig~\ref{fig:wctpr}c, the WCT has indeed a good performance for a noisy signal without clouds. By including clouds in the RCS, was how that at least two local minima appear as in the gradient method (see Fig.~\ref{fig:ideals}d), which difficults the identification of the point value that represents the PBL top. To overcome this, we propose an interplay between the two methods as follows. Pointed out in Section ~\ref{sec:meth}, the gradient method is useful to find both the inflection points and the inflection interval over the idealized profile and experimental RCS. The quality of the results using this method relies on the noise-signal ratio, then, if we have a completely free noise signal we will have only one inflection interval which we define as our threshold.

To obtain a completely free noise signal by experimental RCS, we applied a moving low-pass filter of 20 units of width to attenuate the high frequencies that represent the undesired information; then, we fitted this new smooth signal to Eq.~\ref{eq:simueq} to build an "ideal signal" by our experimental data. At this stage, we obtained the fit parameters to neglect the noise and cloud peaks, leaving a noise-free based on our experimental RCS through a nonlinear least squares technique. Next, by applying the gradient method to the results from the fitting, we get an inflection interval on the entrainment zone where the PBL top position is to be expected; this interval is the threshold obtained by the gradient method. Figures ~\ref{fig:ideals}b and ~\ref{fig:ideals}d show the resulting thresholds (in magenta) with the location of the PBL top. 

The next step to find a suitable approximation to the PBL top is to apply the WCT over the original experimental RCS. To ensure a right selection of WCT parameters, dilation and translation, we took into account an special consideration: Any experimental LiDAR signal retrieved at night and closer to the sunrise and sunset should be transformed by WCT using low values of the $a$ parameter, if compared to any experimental LiDAR signal retrieved around noon, since at noon the well-mixed layer reaches its maximum altitude\cite{compton2013determination}. Therefore, we define a set of values for $a$ and $b$ parameters related to the daytime based on the following intervals:

\begin{equation}
\label{eq:indexa}
\begin{array}{c}
0.15 < a \leq 0.3\\
a > 0.3
\end{array}
\end{equation}

\begin{figure} [ht]
   \begin{center}
   \begin{tabular}{c} 
   \includegraphics[width=12.5cm,height=12.5cm,keepaspectratio]{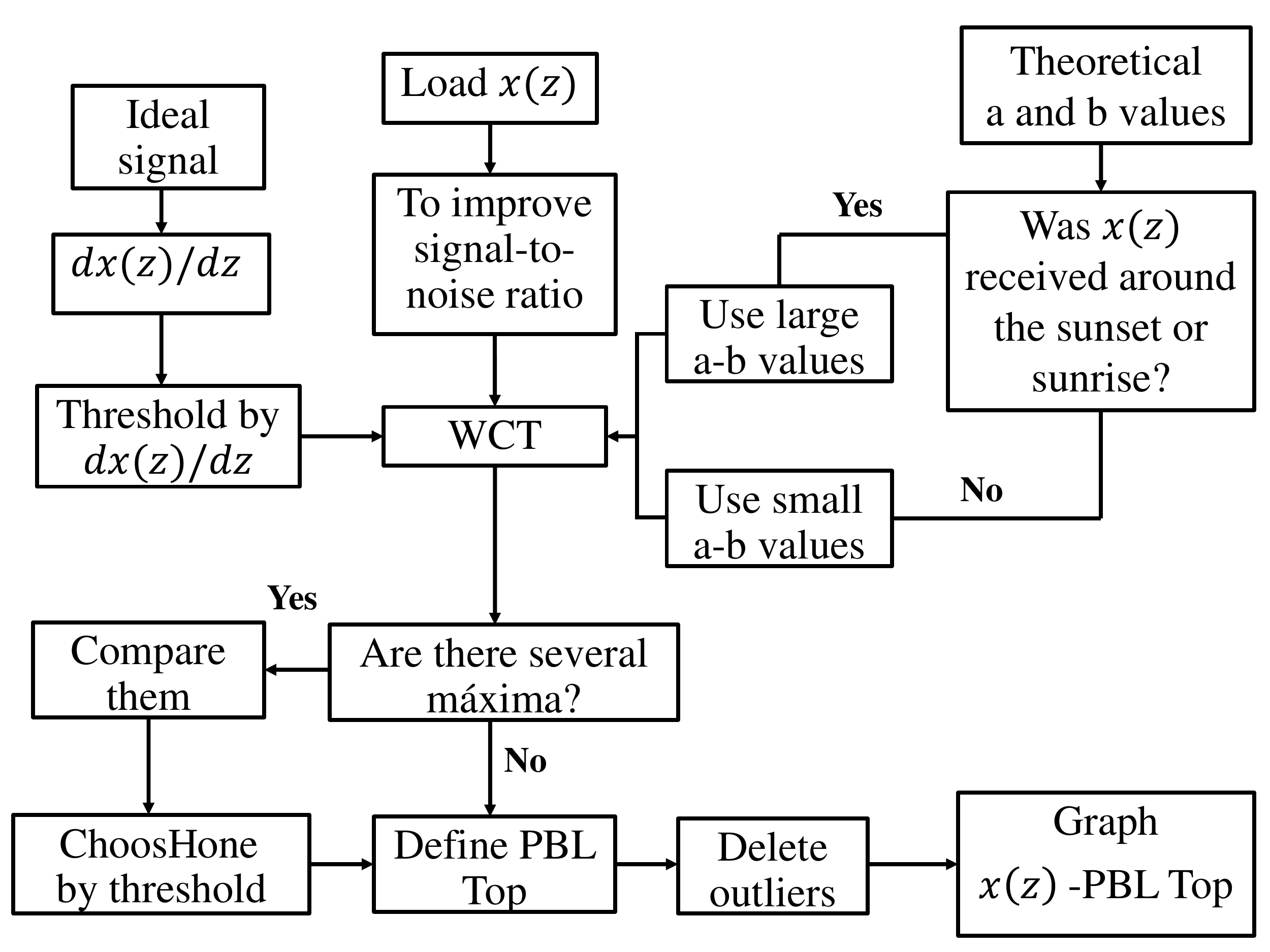}
   \end{tabular}
   \end{center}
   \caption[example] 
   { \label{fig:flow} 
 The algorithm proposed for PBL top detection employs the Gradient method as a threshold for the WCT method.}
   \end{figure}

The first line of Eq.~\ref{eq:indexa} is useful to LiDAR returns closer to the sunrise and sunset, whereas the second inequality is useful when LiDAR returns are closer to the noon. With this suitable set of values for $a$ and inflections intervals from "ideal signals" based on experimental RCS, we can run our algorithm. The novelty of this work is the use of the threshold from the gradient method to iterate the WCT until we arrive at the right value of $a$ from our set, that shows the peak of WCT within the threshold. Finally, the last step consists of debugging the obtained results of local maximum with WCT. By employing a nearest neighbor analysis to evaluate the temporal consistency of our results, and dismiss outliers. The Fig.~\ref{fig:flow} presents the workflow of the algorithm employed in the writing of our numerical code.
   
\section{Results and Discussions}
\label{sec:results}

In this section, we analyze the results about the PBL top obtained from idealized backscatter profiles and the first time experimental RCS in Santiago de Cali. By using the LiDAR-CIBioFI, we have the first approximation to PBL top in the Colombian southwestern.
\subsection{PBL top by simulated signals}
\label{sec:topsimu}

\begin{figure}[ht]
   \begin{center}
   \begin{tabular}{c} 
   \includegraphics[width=16.5cm,height=16.5cm,keepaspectratio]{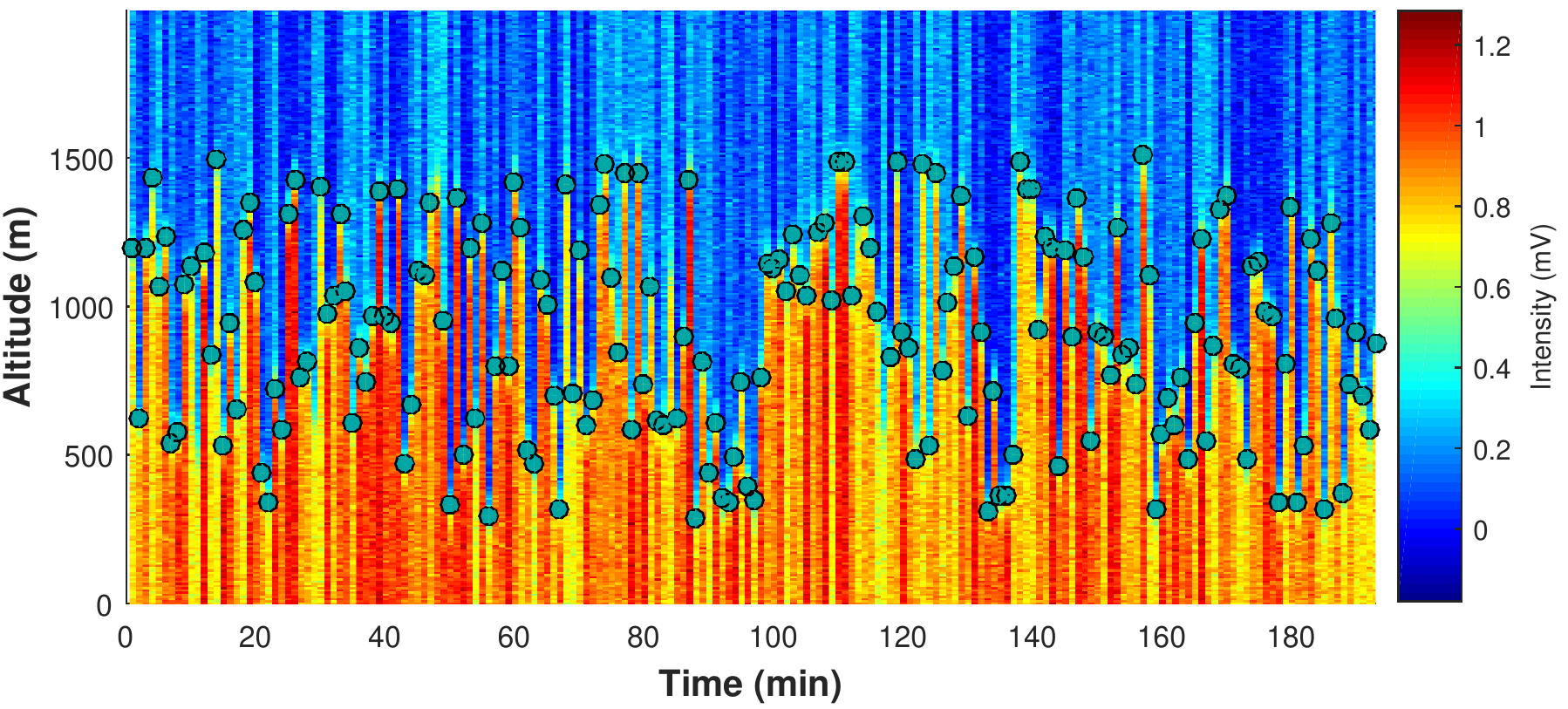}
   \end{tabular}
   \end{center}
   \caption[example] 
   { \label{fig:pblsim} 
Simulated backscatter diagram with random values of $x_{m}$, $x_{u}$, $z_{m}$ and $s$ parameters. For each signal we have ensured large differences between $x_{m}$ and $x_{u}$ to warranty the highlight transition between the well-mixed layer and the free troposphere. The green dots represent the PBL top detected by the algorithm of Fig.~\ref{fig:flow}.}
   \end{figure}

The Fig.~\ref{fig:pblsim} illustrates the backscatter diagram of an ideal backscatter profile. As already mentioned, in Section~\ref{sec:topsimu}, those signals present strong changes of backscattering to detect the PBL top in a well-differentiated mixed layer. The red color symbolizes high values of backscattering and blue colors the free troposphere. The transition between both colors defines the entrainment zone in which the presence of the PBL top is expected. 

The green dots represent the PBL top for each profile, due to the randomly simulated  signals there is not a clear pattern in its temporal behavior. If the green dots avoid the free troposphere or the mixed layer, then we have an indicator of the high algorithm performance. However, the analyzed signals do not account for clouds information, and the evaluation of the methods is incomplete. In the next section, we will examine the performance of the algorithm using an experimental LiDAR signal in presence of clouds.

\begin{figure}[ht]
   \begin{center}
   \begin{tabular}{c} 
   \includegraphics[width=16.2cm,height=16.2cm,keepaspectratio]{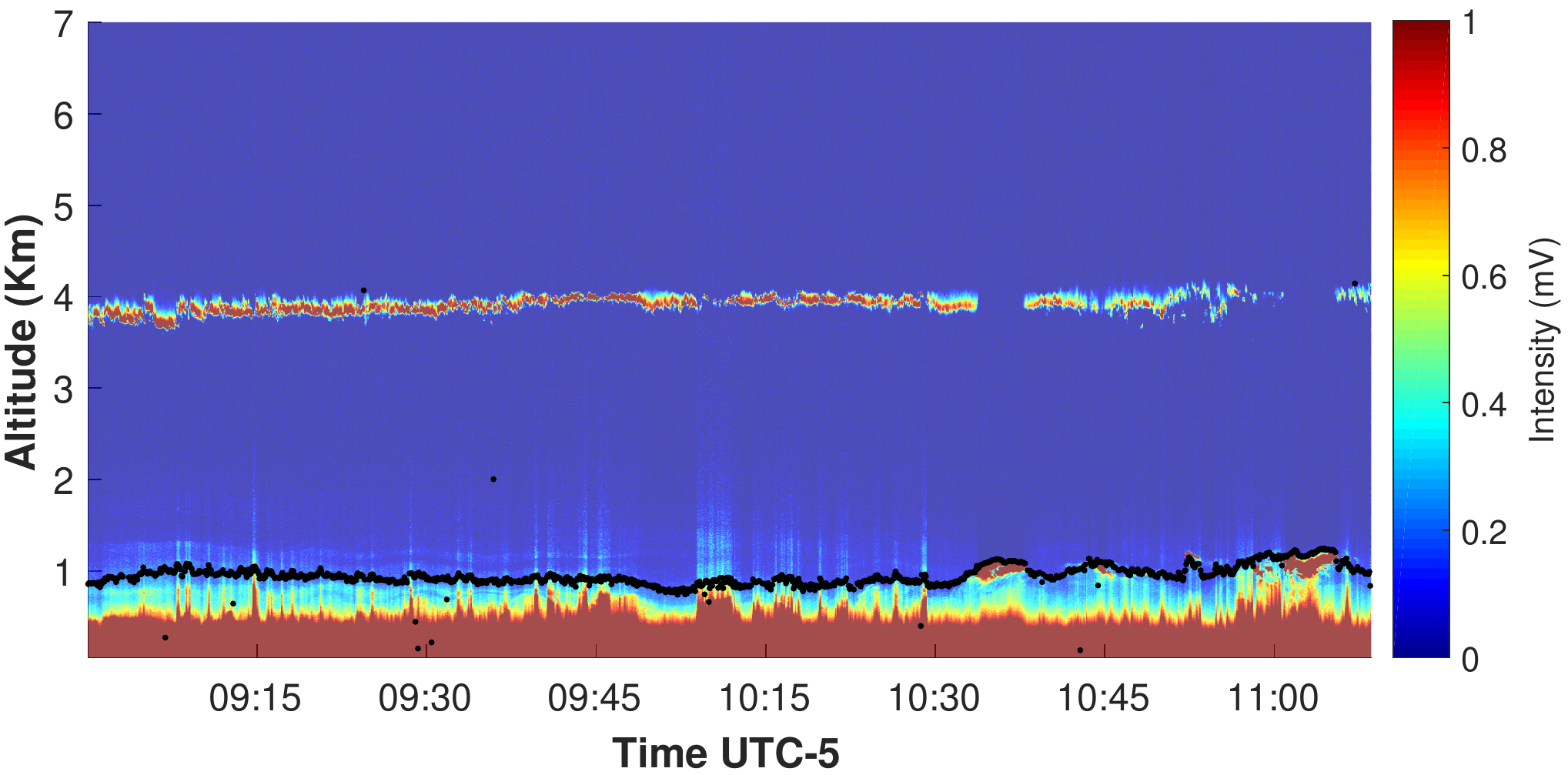}
   \end{tabular}
   \end{center}
   \caption[example] 
   { \label{fig:pblex} 
Experimental backscatter retrieved diagram  with LiDAR-CIBioFI in a coaxial statical configuration. All signals involved in this diagram are composed of 200 pulses with 532 nm of wavelength. This experimental backscatter diagram represents the first PBL observation for the Colombian southwestern.}
\end{figure} 

\subsection{PBL top in experimental signals}
\label{sec:topexp}

Unlike the simulated backscatter diagram, shown in Fig.~\ref{fig:pblsim}, the graph of Fig.~\ref{fig:pblex} is based on experimental signals measured with the LiDAR-CIBioFI system. The Fig.~\ref{fig:pblex} exhibits a set of clouds with values of backscattering similar to the one in the well-mixed layer. Then, two peaks will result after the application of the WCT and the threshold by the Gradient method localizes the black dots inside the entrainment zone. Thus, the protocol to detect the PBL top works, in this case we have a dataset constituted by one hundred (100) experimental signals retrieved each one by two hundred (200) pulses, and only seven indicator dots are out of the expected range. In our experimental detection this amount represents a 93\% of efficiency for the algorithm proposed in Section~\ref{sec:grad} and Fig.~\ref{fig:flow}.

The signals employed in the Fig.~\ref{fig:pblex} were measured on the 23rd of July 2018 at morning in Cali. Here, we observe that in this period the well-mixed layer reaches an average altitude of 0.5 Km, whereas the entrainment zone reaches a maximum average altitude of 1.4 Km, meaning a stable behavior of the PBL. 

The results here reported constitute the first data set of the LiDAR-CIBioFI system as a monitoring station of the air quality and the dynamics of the vertical column atmospheric in the city of Cali. Moreover, it is expected that future observations and research advance towards other modes of analysis such as those due to inelastic scattering. We expected that the new results can be used to validate the performance of the algorithmic proposal here presented and strengthen the understanding of such physical processes.

\section{Acknowledgments}
The authors acknowledge the financial support from CIBioFI, The Colombian Science, Technology and Innovation Fund-General Royalties System (Fondo CTeI-Sistema General de Regal\'ias) and Gobernaci\'on del Valle del Cauca, under contract No. BPIN 2013000100007. J.C. acknowledges Universidad del Valle for partial support (Grant No. CI 71137) and gratefully acknowledges K. Rodr\'iguez, W. Morales, P. Ristori and A. Comer\'on for a critical reading of the manuscript.

\bibliography{report} 
\bibliographystyle{spiebib} 

\end{document}